\pdfoutput=1 
\documentclass{article}
\usepackage{spconf,amsmath,graphicx}
\usepackage{spconf,amsmath,graphicx}
\usepackage{subfigure}
\usepackage{amsmath}
\usepackage{amssymb}
\usepackage{multirow}
\usepackage{booktabs}
\usepackage{enumitem}
\setlist{nosep, leftmargin=14pt}

\usepackage{mwe} 

\title{Single Neuron Segmentation using Graph-based Global Reasoning with Auxiliary Skeleton Loss from 3D Optical Microscope Images}
\name{Heng Wang$^{\star}$, Yang Song$^{\odot}$, Chaoyi Zhang$^{\star}$, Jianhui Yu$^{\star}$, Siqi Liu$^{\diamond}$, Hanchuan Peng$^{\dagger}$, Weidong Cai$^{\star}$}
\address{$^{\star}$ School of Computer Science, University of Sydney, Australia\\
$^{\odot}$ School of Computer Science and Engineering, University of New South Wales, Australia \\
$^{\diamond}$ Digital Services, Digital Technology \& Innovation, Siemens Healthineers, Princeton NJ, USA\\
$^{\dagger}$ SEU-ALLEN Joint Center, Institute for Brain and Intelligence, Southeast University, Nanjing, China\\}
\begin{document}
%
\maketitle
\begin{abstract}
One of the critical steps in improving accurate single neuron reconstruction from three-dimensional (3D) optical microscope images is the neuronal structure segmentation. However, they are always hard to segment due to the lack in quality. Despite a series of attempts to apply convolutional neural networks (CNNs) on this task, noise and disconnected gaps are still challenging to alleviate with the neglect of the non-local features of graph-like tubular neural structures. Hence, we present an end-to-end segmentation network by jointly considering the local appearance and the global geometry traits through graph reasoning and a skeleton-based auxiliary loss. The evaluation results on the Janelia dataset from the BigNeuron project demonstrate that our proposed method exceeds the counterpart algorithms in performance.
\end{abstract}
\section{Introduction}
\label{sec:intro}


3D neuron reconstruction is a vital process in nervous system analysis. It aims to reconstruct the neuron morphology from 3D optical microscopic images for investigation on neuron identity and functionality. It consists of preprocessing, tracing, and postprocessing steps~\cite{liu2018automated}. The preprocessing step is mainly to perform on neuron image segmentation, which is to assign each voxel in the 3D neuron images a label of being either foreground or background. It is challenging due to the noise around the main neuronal structures and the broken thin-branch structures caused by various imaging conditions and the uneven distribution of neuron fluorescent markers~\cite{intro}.

With the successful development of convolutional neural networks (CNNs) on general image processing tasks~\cite{resnet,fcn}, the famous U-Net~\cite{unet} and V-Net~\cite{vnet} have been proposed to enable end-to-end medical image segmentation further advancing the ability of CNNs in neuron image segmentation task~\cite{addedref1,addedref2,isbi19,isbi20}. However, the convolutional kernels in 3D U-Net only convolve around a small regular grid area defined by the kernel sizes. One way to enlarge the receptive field of 3D U-Net to capture more global connectivity in neuron images is to use fusion of convolutional layers with different kernel size as proposed in the 3D MKF-Net~\cite{multiscale}. However, this stacking of convolution operations requires lots of extra parameters, which is inefficient. Hence, irregular convolutional operation connecting long-range feature locations is needed. 

There have been some attempts to apply graph neural networks (GNNs) on regular-grid biomedical image segmentation problems recently. An airway segmentation method~\cite{ugraph} proposed to construct a graph directly from the feature map by considering each feature point as a graph node. However, since the complexity of graph operation is proportional to the number of graph nodes, it can only happen at the last layer where feature maps are relatively small. To reduce the computational requirement, a 2D vessel segmentation approach~\cite{vgn} introduces a semi-regular vertex sampling process where the most potential foreground point from a regular area of feature points is sampled as one graph node. However, this method requires a pretrained segmentation network and the extent of the sampling area requires hyperparameter tunning.

In this paper, by taking 3D U-Net as the backbone, we propose to construct a graph node from the entire feature map rather than a small regular area of it by learning a spatial-wise attention map. In addition, since the number of graph node is flexibly decided, we can perform graph operation on any layers. We also propose a new adaptive compound loss to assist the learning of both local details and global tree-like skeletons. Our method is evaluated on the Janelia dataset from the Gold166 project~\cite{gold166}. We observe that our proposed method outperforms the counterpart algorithms in removing noise and maintaining the overall skeletons of neuronal structures.

\begin{figure*}[!htb]
\centering     
{\includegraphics[width=\textwidth]{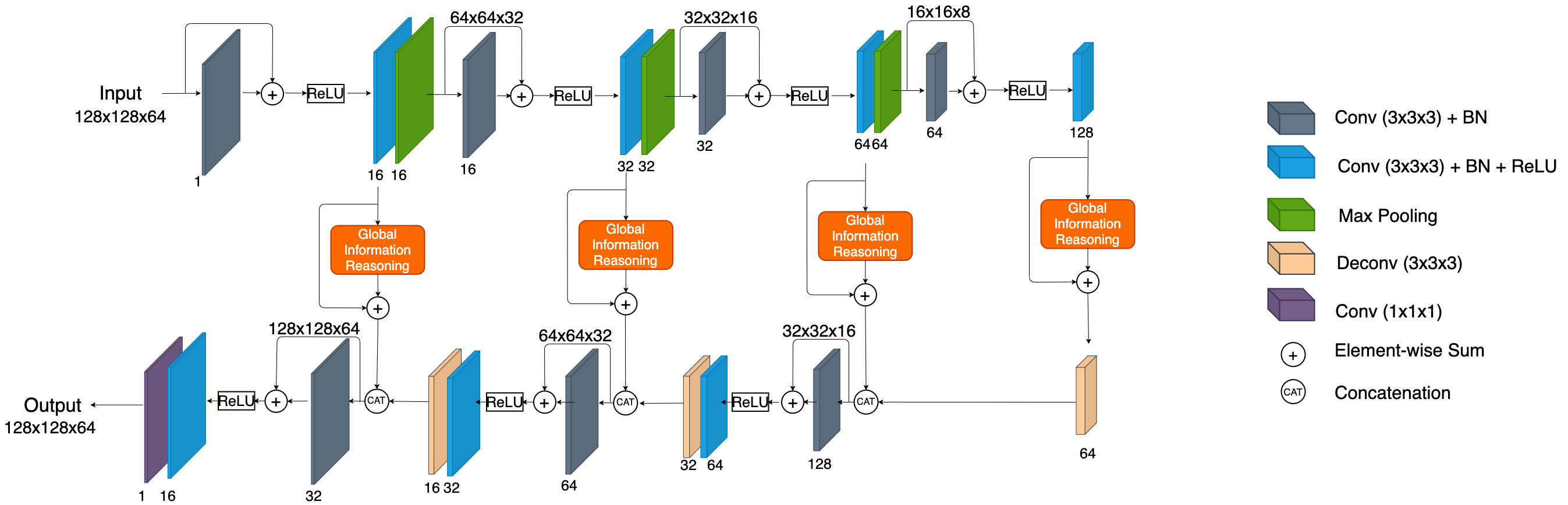}}
\caption{Our proposed network architecture, with input and output channels denoted at the bottom and top of each layer.} 
\label{fig:method:overview}
\end{figure*}

\section{Methods}
\label{sec:method}

\subsection{Overview}
\label{sec:method:overview}
The overall architecture of our proposed network is illustrated in Figure~\ref{fig:method:overview}. The upper part is the encoding path to extract high-order semantic features while the lower part is the decoding path where learned representation is leveraged to full resolution to achieve the end-to-end training. For both the encoding and decoding paths, we utilize skip connections~\cite{resnet} to stabilize the flow of gradient during back-propagation update. The intermediate Global Information Reasoning (GIR) block is proposed to learn complimentary global information by reasoning among the local responses. The learned global features combined with the local representations are then fed into the decoding path to make voxel-wise prediction. More details are elaborated in Section~\ref{sec:method:gir}. We also present a novel adaptive compound loss in Section~\ref{sec:method:loss} to guide the learning on the global graphical neuronal structures while preserving the voxel-wise classification ability.

\subsection{Global Information Reasoning}
\label{sec:method:gir}

\subsubsection{Graph Construction}
\label{sec:sec:method:gnode}
\begin{figure}[!htb]
\centering 
\includegraphics[width=0.45\textwidth]{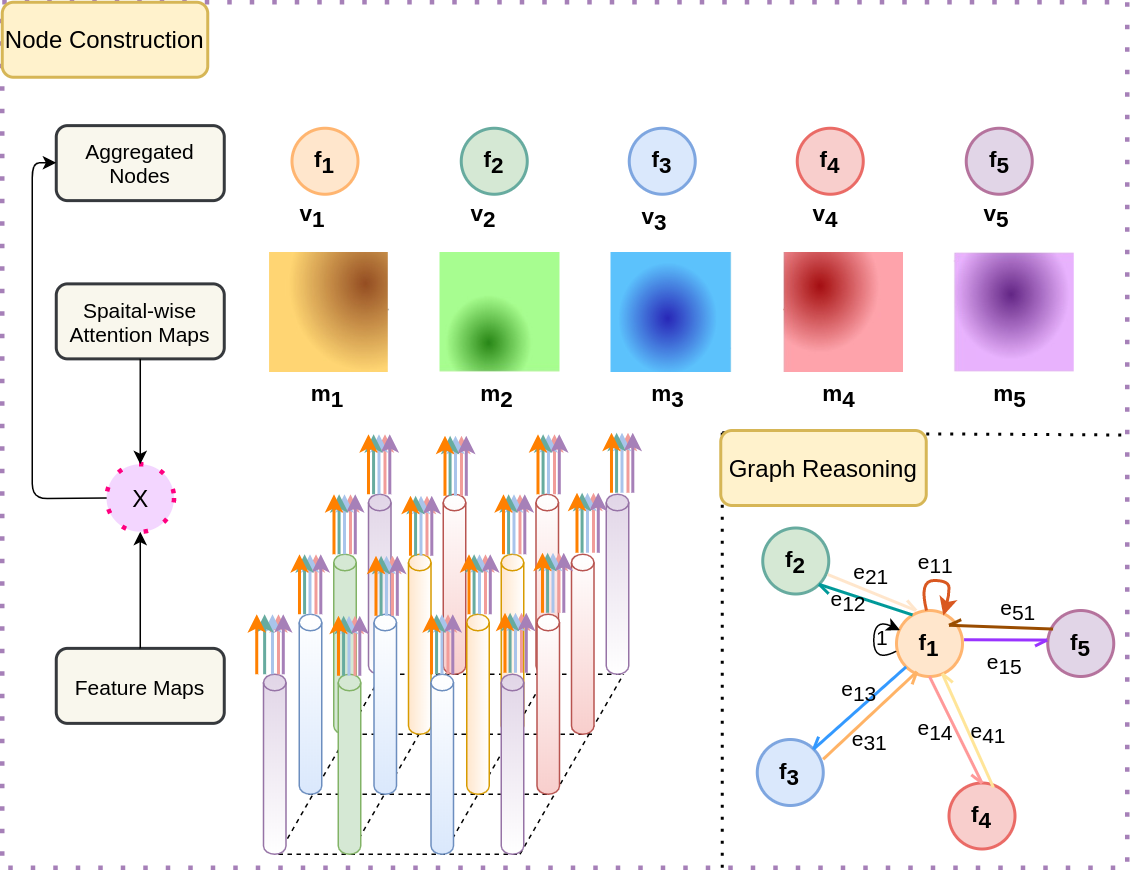}
\caption{Illustration of the graph node construction and the graph reasoning aggregation process for node $v_1$. Feature map is colored in accordance with the nodes they make the most contribution to, while \textbf{x} denotes weighted summation.}
\label{fig:node}
\end{figure}
In order to perform irregular graph reasoning on the regular grid neuron feature data, we first design a method to construct a graph $\mathcal{G}: (\mathcal{V},\mathcal{E})$ from the input $X\in \mathbb{R}^{S \times C_{in}}$ inside the GIR block. Here, $\mathcal{V}$ represents the set of graph nodes $v_{i:1, 2, ..., N}$ and $\mathcal{E}$ is the set of edges. $S=H\times W\times D$ refers to the spatial dimension of the input $X$ with $H$, $W$, and $D$ denoting the height, the width, and the depth of it, and $C_{in}$ is the channel dimension. To save the computation cost, we first group these $S$ input points $x_i$ into $N$ superpoints (graph nodes $v_i$). To make each superpoint contain global information with long-range connectivity, we propose to use $N$ learnable spatial-wise attention maps $M\in \mathbb{R}^{N \times S}$ to define how each point is mapped to these superpoints. Concretely, each attention map $m_i\in \mathbb{R}^{1\times S}$ encodes the relation between $S$ input points and one unique superpoint $v_i$. Inspired by the works~\cite{glore,graphonomy,glore3,glore2}, we use a $1\times 1\times 1$ unit convolution with $C_{in}$ input dimension and $N$ output dimension to learn such attention maps $M$ from $X$ based on the intrinsic feature patterns of the input. Then the initial state of the graph nodes $F\in \mathbb{R}^{N\times C_{gcn}}$ is formulated as $F =M g(X)$ where $g(\cdot)$ refers to another unit convolution operation with $C_{in}$ input dimension and $C_{gcn}$ output dimension. The aim of $g(\cdot)$ is to generate more compact graphical features for later reasoning. An example of such a node construction process is also illustrated in Figure~\ref{fig:node} with input feature data being mapped to five graph nodes. Five spatial-wise attention maps $m_{i:1,2,...,5}\in \mathbb{R}^{1\times S}$ are learned from the feature maps and then the weighted summation of all feature vectors $x_{s}\in \mathbb{R}^{1\times C_{gcn}}$ is computed to get the corresponding graph node $v_i$. 

Since we already reduce the number of graph nodes using grouping and $N$ could be a small number, we can introduce edges between every pair of graph nodes to do full reasoning, which captures the long-range connectivity among all the superpoints. We use an adjacency map $A\in \mathbb{R}^{N \times N}$ to represent how graph nodes aggregate information from each other. For example, $A_{i, j}$ denotes how node $v_i$ influences node $v_j$. 




\subsubsection{Graph Reasoning}
\label{sec:sec:method:greason}
The main purpose of graph reasoning here is to learn the global features. It updates the state from $F\in \mathbb{R}^{N\times C_{gcn}}$ to $F\in \mathbb{R}^{N\times C^{\prime}_{gcn}}$ by aggregating information from the neighbourhood allowing interaction between long-range points. Here, we use the graph convolution operation~\cite{gcn} to perform such a global-level reasoning. \newline

\noindent \textbf{Aggregation.}
 The first step is to aggregate the feature (i.e. the node state $f_i$) of each node $v_i$ from all the graph nodes including itself based on the edge weight matrix $A$, which is randomly initialized as $N\times N$ trainable parameters. With the iterative aggregation process, $A$ is updated accordingly through gradient descent. 
The activated aggregated node feature $f^{agg}_i$ from the initial state $f_i$ through $A$ for one node $v_i$ is formulated as follows:
\begin{equation}
    f^{agg}_i = \sigma(f_i + \sum_j^NA(j, i)f_j),
\end{equation}
where $\sigma(\cdot)$ refers to the rectified linear unit activation function. The right corner of Figure~\ref{fig:node} displays one example about how the information from a node $v_1$'s neighbourhood is aggregated. Note that an additional self-loop is introduced here to emphasize the feature of the node state $f_1$ itself. \newline

\noindent \textbf{Transformation.}
Once the information is fused for all graph nodes, we extract global features $F^{out} = F^{agg}  W$ from aggregated node features $F^{agg}\in \mathbb{R}^{N\times C_{gcn}}$ where $W\in \mathbb{R}^{C_{gcn}\times C^{\prime}_{gcn}}$ is the trainable parameters to transform the feature dimension from $C_{gcn}$ to $C^{\prime}_{gcn}$. The spatial-wise attention map $M$ is needed again to map the node features back to the spatial feature space as follows:
\begin{equation}
    X' =h(M^T F^{out}), \\
\end{equation}
where $h(\cdot)$ refers to another unit convolution operation with $C^{\prime}_{gcn}$ input dimension and $C_{in}$ output dimension to align with the local feature $X$ along the channel dimension. A batch normalization layer is then attached before the output features are combined with the local feature maps.

\subsection{Skeleton-aware Adaptive Compound Loss}
\label{sec:method:loss}
In the task of 3D neuron segmentation, we normally depend on binary cross entropy loss ($L_{CE}$) to simulate the distribution of the ground truth segmentation. However, $L_{CE}$ does not consider the tubular connectivity of the neuron arborization. To enforce the graphical structure into the learning process, we propose to incorporate an auxiliary loss named skeletonLoss ($L_{SKL}$) into the existing $L_{CE}$. As the name suggests, $L_{SKL}$ is based on the comparison between the skeleton of the prediction and that of the ground truth segmentation. Inspired by the recently proposed clDiceLoss~\cite{cldice}, we use the combination of min and max 3D pooling to generate the skeleton on the fly during the training so that the loss can be backpropagated through gradient descent. In details, the $L_{SKL}$ is defined as follows:
\begin{equation}
\begin{split}
& p = \frac{skl_p\cdot skl_l+\delta}{skl_p+\delta},\quad r = \frac{skl_p\cdot skl_l+\delta}{skl_l+\delta}, \\
& L_{SKL} = 1.0-\frac{2\cdot p\cdot r}{p+r},\\
\end{split}
\end{equation}
where $skl_p$ is the skeleton for the prediction after sigmoid activation function and $skl_l$ is the skeleton for the ground truth segmentation. $\delta$ is the smooth value and set to 1. To stabilize the training, we control the weight of $L_{CE}$ and $L_{SKL}$ using an adaptive rule according to the progress of the training, and our final objective function now becomes as follows:
\begin{equation}\label{prf}
L = \alpha \cdot L_{CE} + (1-\alpha) \cdot L_{SKL}, \:\alpha = \frac{2}{(1 + e^{-10 * p})} - 1,\\
\end{equation}
where $p$ is the ratio between current iteration and the total number of iterations. Note that the scheduling rule of $\alpha$ is gradually increasing from 0 to 1~\cite{rev} to stress more on the tubular shape at the early training stage.

\begin{figure*}[!t]
\centering     
\subfigure[Original image]{\includegraphics[width=0.18\textwidth]{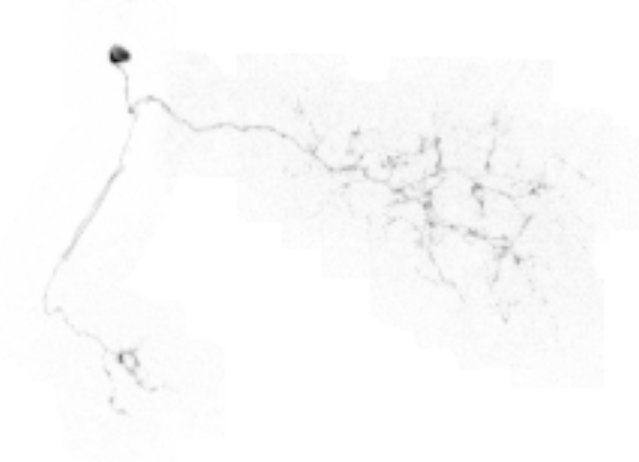}}
\subfigure[Ground truth]{\includegraphics[width=0.18\textwidth]{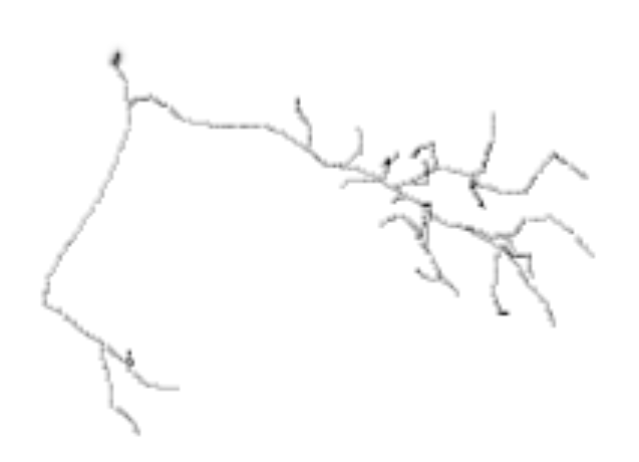}}
\subfigure[3D Res-U-Net]{\includegraphics[width=0.18\textwidth]{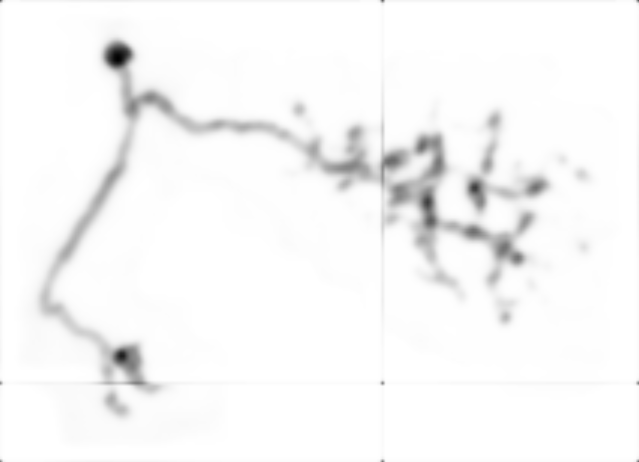}}
\subfigure[3D Res-MKF-Net]{\includegraphics[width=0.18\textwidth]{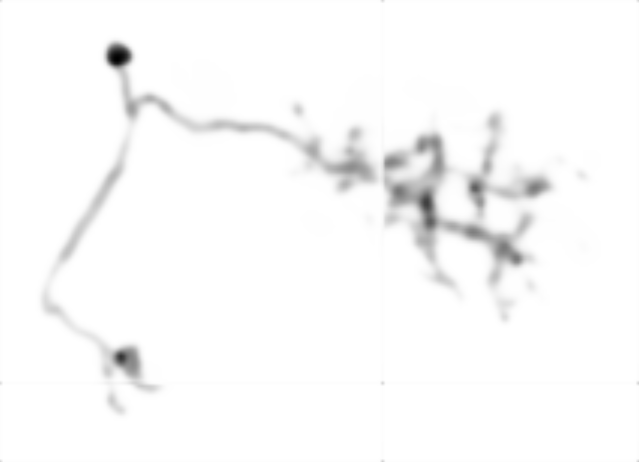}}
\subfigure[Proposed]{\includegraphics[width=0.18\textwidth]{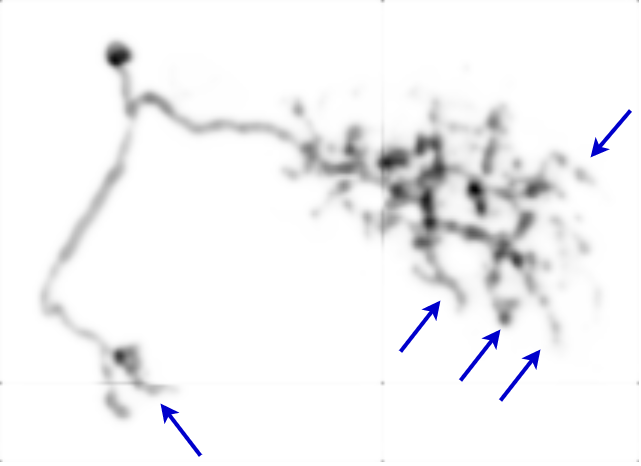}}
\caption{Visualization of the predictions for testing image \textit{timg4} from the baseline 3D Res-U-Net and our proposed method. We project the 3D outputs into 2D plane for better visual comparison.  Some of the structures segmented by the proposed method but not by the baseline are highlighted using blue arrows.}
\label{fig:visual_compare}
\end{figure*}

\section{Experiments and Results}
\label{sec:experiment}

\subsection{Dataset and Implementation Details}
\label{sec:experiment:imp}
We performed the evaluation on 42 3D neuron images from the Janelia dataset of the BigNeuron project. They are randomly divided into 35, 3, and 4 samples for training, validation, and testing, respectively, whose average image size is $197\times199\times168$, $110\times256\times112$, and $262\times159\times181$, respectively. Random horizontal and vertical flipping, rotation, and cropping with the patch size $128\times128\times64$ are adopted to enlarge the training set.

All the models were reproduced from scratch with the Adam optimizer~\cite{adam}. The learning rate was set to $1\times 10^{-3}$ and the weight decay was $5\times 10^{-4}$, with batch size set to 8 and early stopping technique adopted. For models with the GIR block, $N$, $C_{gcn}$, and $C^{\prime}_{gcn}$ are set to $\frac{C_{in}}{4}$, $\frac{C_{in}}{2}$, and $\frac{C_{in}}{2}$, respectively, as suggested by ~\cite{glore}. We applied a Gaussian filter ($\sigma$=0.8) on all sets when using the proposed adaptive compound loss. If the training epoch is less than 200, the total number of iterations is computed for 200 epochs. Once the training lasts over 200 epochs, $p$ is changed to two times the ratio between the current iteration and total iterations for 300 epochs. 
\begin{table}[!b]

\begin{center}
\resizebox{\linewidth}{!}{%
\begin{tabular}{|l|l|l|l|l|l|l|l|}
\hline
Method & $\#$Params (M)& Metric  & timg1         & timg2             & timg3  & timg4         & Overall Results   \\
\hline
\multirow{3}{*}{Simple}& \multirow{3}{*}{$-$}& 
Precision &  $0.2368$ & $0.4012$ & $0.5335$ & $0.2393$ & $0.3527$ $\boldsymbol{\pm}$ $0.1430$   \\
& & Recall& $0.4077$ & $0.5491$ & $0.6754$ & $0.2600$ & $0.4731$ $\boldsymbol{\pm}$ $0.1793$ \\
& & F1 & $0.2995$ & $0.4636$ & $0.5961$ & $0.2492$ & $0.4021$ $\boldsymbol{\pm}$ $0.1585$\\
\hline
\multirow{3}{*}{3D Res-U-Net} & \multirow{3}{*}{$1.40$}& 
 Precision & $0.2977$ & $0.4731$ & $0.6019$ & $0.2922$ & $0.4162$ $\boldsymbol{\pm}$ $0.1496$\\
 & & Recall& $0.5296$ & $0.6907$ & $0.7482$ & $0.5173$ & $0.6215$ $\boldsymbol{\pm}$ $0.1157$ \\
 & & F1& $0.3812$ & $0.5616$ & $0.6671$ & $0.3735$ & $0.4958$ $\boldsymbol{\pm}$ $0.1435$\\
\hline
\multirow{3}{*}{3D Res-MKF-Net} & \multirow{3}{*}{$1.55$ ($+11\%$)}&
 Precision & $0.3200$ & $0.5101$ & $0.5876$ & $0.3227$ & $0.4351$ $\boldsymbol{\pm}$ $0.1351$\\
 & & Recall& $0.4933$ & $0.6691$ & $0.7549$ & $0.4704$ & $0.5969$ $\boldsymbol{\pm}$ $0.1377$\\
 & & F1& $0.3882$ & $0.5789$ & $ 0.6608$ & $0.3828$ & $0.5027$ $\boldsymbol{\pm}$ $0.1394$\\
\hline
\multirow{3}{*}{Proposed} & \multirow{3}{*}{$1.43$ ($+2\%$)}&
 Precision & $0.3371$ & $0.5156$ & $0.6007$ & $0.3056$ & $\textbf{0.4398}$ $\boldsymbol{\pm}$ $0.1416$\\
 & & Recall& $0.5358$ & $0.7256$ & $0.7626$ & $0.5177$ & $\textbf{0.6354}$ $\boldsymbol{\pm}$ $0.1266$\\
 & & F1& $0.4138$ & $0.6028$ & $ 0.6720$ & $0.3844$ & $\textbf{0.5183}$ $\boldsymbol{\pm}$ $0.1410$\\
\hline
\end{tabular}}
\caption{The quantitative segmentation comparison among the conventional thresholding method, 3D Res-U-Net, 3D Res-MKF-Net, and our proposed network on four testing images. Our number of increased parameters is largely less than that of the 3D Res-MKF-Net, compared to 3D Res-U-Net.}
\label{tab:qcompare}
\end{center}
\end{table}

\subsection{Results and Discussion}
\label{sec:experiment:result}
We first present our result comparison with the baseline 3D U-Net~\cite{unet}, 3D multiscale kernel fusion network (MKF-Net)~\cite{multiscale}, and a simple traditional thresholding method where only a value is applied on the original images to filter out noise. We add the same skip connection to both U-Net and MKF-Net as our proposed network. We compute F1-score for different probability thresholds for each model on each testing image. The best F1-score as well as the corresponding precision and recall values are listed as the evaluation metrics in Table~\ref{tab:qcompare}. According to the quantitative results shown in Table~\ref{tab:qcompare}, our proposed method performs the best on all these three metrics. It demonstrates that the global information is necessary when distinguishing between the neuronal structures and the background. The graph-based global information reasoning is better than the stacking of several convolutional layers as our proposed network exceeds the 3D Res-MKF-Net. The 3D Res-U-Net outperforms the simple thresholding method by over $6\%$, $14\%$, and $9\%$ respectively on the three metrics while our proposed method further exceeds the 3D Res-U-Net by over $2\%$, $1\%$, and $2\%$ respectively.

We visualize the prediction probability results from different methods in Figure~\ref{fig:visual_compare}. Note that the 3D probability results are projected to 2D plane on purpose in order to see the skeleton clearly. Our proposed method is able to make a more complete segmentation compared to the other two algorithms. It is reasonable since we incorporate more global dependency in our learning process, and the segmentation of overall skeleton is guided by the auxiliary loss.

Extensive ablation studies are presented to verify the contribution made by each proposed component towards the final segmentation results, and report averaged F1-score in Table~\ref{tab:ablation}. As indicated by result of model B, the incorporation of the proposed global information reasoning module is able to improve the performance. This demonstrates that the global information is beneficial to the neuronal structure segmentation task. Model D exceeds model C by over $1\%$ indicating that the proposed adaptive compound loss improves the capturing of the graphical tubular shape.
\begin{table}[!t]

\begin{center}
\resizebox{0.9\linewidth}{!}{%
\begin{tabular}{c|cccc|c}
\toprule
Model &  $L_{CE}$ & GIR      & Gaussian Filtering             & Adaptive Compound Loss & Best F1-Score\\
\hline
A& \checkmark  &                     &    &   & $0.4949$     \\
\hline
B & \checkmark &   \checkmark    &   &    & $0.5054$\\
\hline
C & \checkmark  & \checkmark     & \checkmark &  & $0.5004$\\
\hline
D   &   &  \checkmark  & \checkmark  & \checkmark & $0.5157$\\
\bottomrule
\end{tabular}}
\caption{Ablation studies of our proposed method. }
\label{tab:ablation}
\end{center}
\end{table}
\section{Conclusion}
\label{sec:conclusion}
To improve the 3D neuron image segmentation ability, we propose a graph-based global information reasoning module to utilize the global connectivity among the graphical neuronal structures in 3D optical microscopic neuron images. The proposed module constructs a fully-connected graph based on spatial-wise attention maps and local responses. It then utilizes long-range connections to generate global responses by aggregating neighbourhood information while the relationship between each node is being learned. We also present a novel adaptive compound loss to take the tubular shape of the neuron arborization into consideration for a further improvement. Our evaluation on the Janelia dataset demonstrates that the proposed method is able to extract global connectivity and preserve the global geometry traits of the neuronal structures and it outperforms the counterpart algorithms. Ablation studies show the contribution of each component towards 3D neuron segmentation performance.

\section{Compliance with Ethical Standards}
\label{sec:ethics}
This research study was conducted retrospectively using fly neuron data made available in open access by~\cite{gold166}. Ethical approval was not required as confirmed by the license attached with the open access data.

\section{Acknowledgments}
\label{sec:acknowledgments}
No funding was received for conducting this study. The authors have no relevant financial or non-financial interests to disclose. 


\bibliographystyle{IEEEbib}

\end{document}